\documentclass[cameraready]{Interspeech}
\usepackage{booktabs}
\usepackage{multirow}

\title{SETEAB: Multiscale approach with Squeeze-and-Excitation Temporal Enhanced Aware Block for Speech Emotion Recognition}

\author[affiliation={1,2}, orcid=0009-0006-9774-594X]{Duy}{Vo}
\author[affiliation={1,2, 3}, orcid=0000-0001-8438-6585]{Kiet Anh}{Hoang}
\author[affiliation={1,2}, orcid=0000-0002-9014-1506, correspondingauthor]{Hao}{Do}

\address{
    $^1$ Faculty of Information Technology, University of Science, Ho Chi Minh City, Vietnam \\
    $^2$ Vietnam National University, Ho Chi Minh City, Vietnam \\
    $^3$ UNEY, Switzerland 
}

\email{\{24C11006,22C15011\}@student.hcmus.edu.vn, ddhao@fit.hcmus.edu.vn}


\usepackage{comment}

\begin{document}

\maketitle

\begin{abstract}
   This paper proposes a novel lightweight multiscale architecture for speech emotion recognition (SER) with three key innovations. First, a depthwise convolution-based subsampling module is introduced to reduce model size and computation while preserving salient emotional cues. Second, a Squeeze-and-Excitation block is integrated to enhance channel-wise recalibration and improve representation robustness. Third, a new Temporal Enhanced Aware Block is designed to strengthen temporal dependency modeling and produce more discriminative emotion-aware features. The proposed model is explicitly designed to jointly improve compactness, recognition performance, and generalizability. Experiments on benchmark SER datasets show that our method achieves higher accuracy with reduced computational complexity, while also delivering stronger cross-corpus performance than most recent advanced networks for SER.
\end{abstract}

\section{Introduction}
\label{sec:intro}

Speech emotion recognition (SER) aims to identify human emotional states from speech signals, enabling more natural and emotionally aware human-computer interaction~\cite{schuller2018speech}. It is essential for applications such as healthcare monitoring, adaptive customer service, in-vehicle safety, and call-center analytics~\cite{anagnostopoulos2015features}. Despite substantial advances in speech processing, robust emotion understanding remains challenging for practical intelligent systems due to the inherent variability in speaking styles, linguistic backgrounds, and environmental noise.

Early SER systems relied on handcrafted acoustic features combined with conventional classifiers such as support vector machines~\cite{tuncer2021automated}. With the rise of deep learning, end-to-end neural architectures have largely replaced manual feature engineering. Convolutional neural networks effectively capture local spectral patterns~\cite{ilyas2021pseudo, wen2021application}, while recurrent neural networks model temporal dependencies across speech frames~\cite{rajamani2021novel, wang2020speech}. More recently, self-supervised learning models such as Wav2Vec~2.0~\cite{baevski2020wav2vec} and WavLM~\cite{chen2022wavlm} have achieved state-of-the-art performance~\cite{pepino2021emotion, ma2024emotion2vec, diatlova24_odyssey}. However, their high computational cost often limits deployment in resource-constrained edge environments.


To better balance accuracy and efficiency, lightweight SER architectures have attracted increasing attention. Approaches based on depthwise separable convolutions~\cite{zhong2020lightweight} and fully convolutional designs such as Light-SERNet~\cite{aftab2022light} significantly reduce parameter counts while maintaining competitive performance. Among efficient temporal modeling frameworks, TIM-Net~\cite{ye2023temporal} has emerged as a strong baseline, employing Temporal-Aware Blocks (TABs) with dilated causal convolutions to capture multi-scale temporal dependencies. Building upon this architecture, MS-SENet~\cite{10447232} further enhances TIM-Net by incorporating multi-scale feature fusion with Squeeze-and-Excitation mechanisms to improve emotional representation learning.

Despite its effectiveness, the standard TAB architecture has several limitations: the absence of explicit normalization may cause optimization instability and gradient attenuation; standard convolutions can introduce feature redundancy and limited temporal-resolution control; and direct bidirectional summation assumes equal past-future contributions, which may not reflect emotional dynamics accurately.

To address these issues, we propose a multiscale speech emotion recognition framework based on the \textbf{Squeeze-and-Excitation Temporal Enhanced Aware Block (SETEAB)}. Extensive experiments on benchmark SER datasets demonstrate that our method achieves an outstanding balance among accuracy, computational efficiency, and cross-corpus generalization. The main contributions are:

\begin{itemize}
\item We introduce a novel \textbf{Temporal Enhanced Aware Block (TEAB)} as an upgrade to the standard TAB to model temporal emotional dynamics more effectively and generate discriminative emotion-aware representations.
\item We develop a \textbf{dynamic fusion strategy} that adaptively integrates forward and backward temporal information by learning their relative importance.
\item We incorporate a \textbf{Squeeze-and-Excitation Res2Block} to enhance multi-scale temporal feature learning and channel-wise recalibration, improving robustness and generalization.
\item We employ \textbf{depthwise convolution subsampling} to reduce computational cost and model size for faster inference, while preserving salient affective cues.
\end{itemize}
\begin{figure*}[t]
    \centering
    \includegraphics[width=\linewidth]{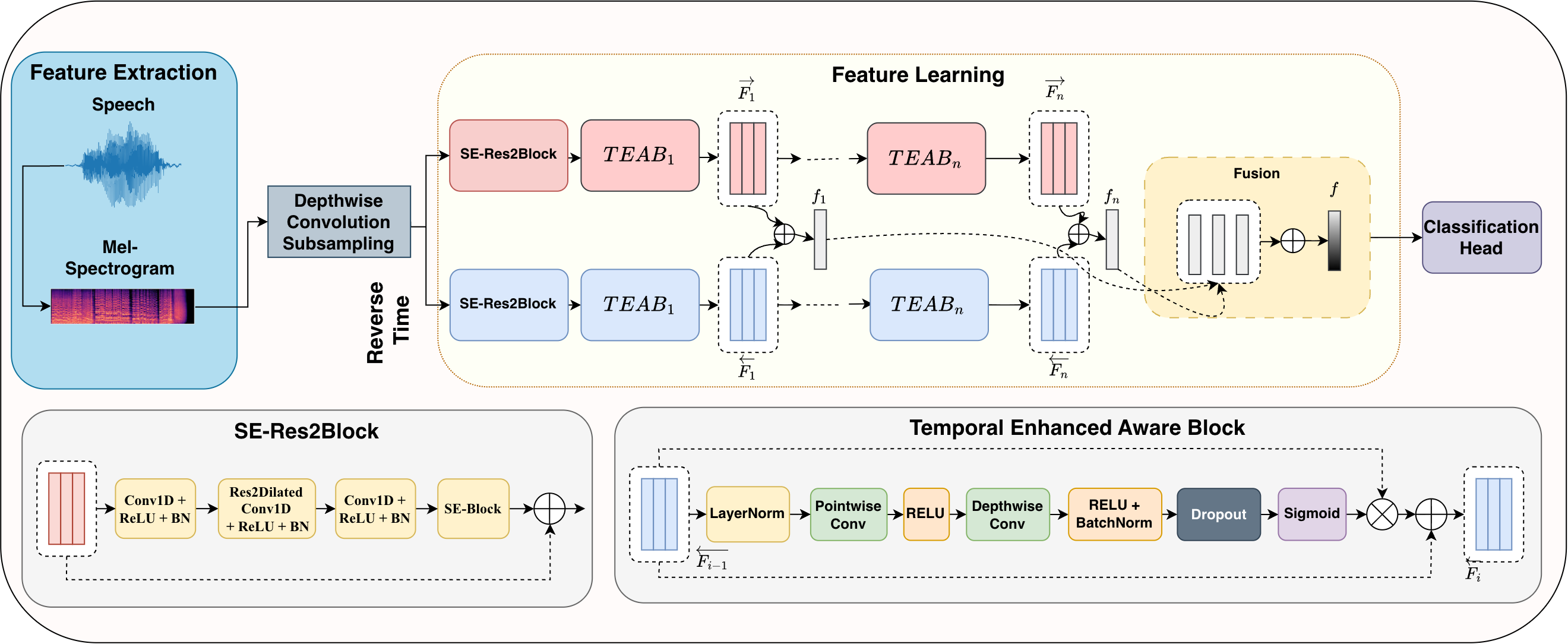}
    \caption{Overview of the proposed \textbf{SETEAB} for speech emotion recognition with four key improvements: depthwise convolution subsampling, SE-Res2Block, the proposed Temporal Enhanced Aware Block (TEAB), and weighted bidirectional fusion (BiF).}
    \label{fig:teab}
\end{figure*}

\section{Method}
\label{sec:method}
The proposed \textbf{SETEAB} framework, shown in Fig.~\ref{fig:teab}, extends TIM-Net with four modifications: \emph{(i)} a \textbf{depthwise convolution subsampling} front-end for early temporal compression, \emph{(ii)} an \textbf{SE-Res2Block} for local multi-scale and channel-aware refinement, \emph{(iii)} a stack of \textbf{Temporal Enhanced Aware Blocks (TEABs)} for bidirectional temporal modeling, and \emph{(iv)} a \textbf{weighted bidirectional fusion} module for adaptive integration of forward and reverse temporal information. Overall, the front-end first removes redundant short-term variation, the SE-Res2Block strengthens local discriminative cues, the TEAB stack progressively refines temporal representations, and the fusion module aggregates directional and multi-level information for emotion prediction.

\subsection{Overall Framework}
\label{subsec:overall}

Given an input utterance $x$, we first extract its Mel-spectrogram:
\begin{equation}
\mathbf{X}=\mathrm{MelSpec}(x), 
\qquad 
\mathbf{X}\in\mathbb{R}^{T\times M},
\end{equation}
where $T$ and $M$ denote the numbers of time frames and Mel bins, respectively. The overall computation is defined as:
\begin{equation}
\mathbf{Z}=\mathcal{D}(\mathbf{X}),
\label{eq:overall_d}
\end{equation}
\begin{equation}
\overrightarrow{\mathbf{F}}_{0}=\mathcal{R}(\mathbf{Z}), 
\qquad
\overleftarrow{\mathbf{F}}_{0}=\mathcal{R}(\mathrm{Rev}(\mathbf{Z})),
\label{eq:overall_r}
\end{equation}
\begin{equation}
\overrightarrow{\mathbf{F}}_{i}=\mathcal{T}_{i}(\overrightarrow{\mathbf{F}}_{i-1}), 
\qquad
\overleftarrow{\mathbf{F}}_{i}=\mathcal{T}_{i}(\overleftarrow{\mathbf{F}}_{i-1}),
\label{eq:overall_t}
\end{equation}
\begin{equation}
\mathbf{f}_{i}=\mathcal{F}_{i}\big(\overrightarrow{\mathbf{F}}_{i},\overleftarrow{\mathbf{F}}_{i}\big), 
\qquad i=1,\dots,n,
\label{eq:overall_f}
\end{equation}
\begin{equation}
\mathbf{f}=\mathcal{A}(\mathbf{f}_{1},\ldots,\mathbf{f}_{n}),
\label{eq:overall_a}
\end{equation}
where $\mathcal{D}(\cdot)$ denotes depthwise subsampling, $\mathcal{R}(\cdot)$ denotes the SE-Res2Block, $\mathcal{T}_{i}(\cdot)$ denotes the $i$-th TEAB, $\mathcal{F}_{i}(\cdot)$ denotes weighted bidirectional fusion, and $\mathcal{A}(\cdot)$ denotes multi-level aggregation. The final representation $\mathbf{f}$ is fed to the classification head.





\subsection{Depthwise Convolution Subsampling}
\label{subsec:dcs}

To reduce temporal redundancy and computational cost at an early stage, we adopt a \textbf{depthwise convolution subsampling} front-end similar to FastConformer~\cite{10389701}. The module achieves a subsampling rate $R=2^{N_S}$ using $N_S$ convolutional blocks, each with stride $2$ and kernel size $3$. The first block applies a standard 2D convolution, while the remaining $N_S-1$ blocks use depthwise separable convolutions:
\begin{align}
\mathbf{S}_1 &= \delta(\mathrm{Conv2D}(\mathbf{X})),\\
\mathbf{S}_i &= \delta(\mathrm{PWConv}(\mathrm{DWConv}(\mathbf{S}_{i-1}))),\quad i=2,\dots,N_S,
\label{eq:dcs_ops}
\end{align}
where $\delta(\cdot)$ denotes the ReLU activation. The temporal length after stage $i$ is calculated as:
\begin{equation}
T_i=\left\lfloor\frac{T_{i-1}-3}{2}\right\rfloor+1,\quad T_0=T,
\label{eq:dcs_len}
\end{equation}
The final feature $\mathbf{Z} = \mathbf{S}_{N_S}$ with length $T' = T_{N_S}$ effectively compresses redundant short-term variations while preserving salient affective information for subsequent temporal modeling.

\subsection{SE-Res2Block}
\label{subsec:seres2}

The subsampled feature is then refined by an \textbf{SE-Res2Block}~\cite{desplanques2020ecapa}, to enhance local multi-scale patterns and channel-wise discriminability before temporal modeling. Starting from $\mathbf{Z}$, the block applies:
\begin{equation}
\mathbf{U}_{1}=\phi_{1}(\mathbf{Z}), \qquad
\mathbf{U}_{2}=\phi_{2}(\mathbf{U}_{1}), \qquad
\mathbf{U}_{3}=\phi_{3}(\mathbf{U}_{2}),
\label{eq:seres2_main}
\end{equation}
where $\phi_{1}$ and $\phi_{3}$ denote Conv1D-ReLU-BN, while $\phi_{2}$ denotes Res2DilatedConv1D-ReLU-BN. Channel-wise recalibration is then performed by:
\begin{equation}
\mathbf{s}
= \sigma\left(
\mathbf{W}_{2}\,
\delta\left(
\mathbf{W}_{1}\,
\mathrm{GAP}(\mathbf{U}_{3})
\right)\right),
\label{eq:seres2_se}
\end{equation}
where $\mathrm{GAP}(\cdot)$ denotes Global Average Pooling. The block output is:
\begin{equation}
\mathcal{R}(\mathbf{Z})
=
\mathbf{Z}
+
\mathbf{s}\odot\mathbf{U}_{3},
\label{eq:seres2_out}
\end{equation}
where $\odot$ denotes element-wise multiplication. This design preserves the original representation while emphasizing emotion-relevant channels.

\subsection{Temporal Enhanced Aware Block}
\label{subsec:teab}

The proposed \textbf{Temporal Enhanced Aware Block (TEAB)} is the core temporal unit of SETEAB. Compared with the TAB in TIM-Net, the TEAB improves stability and feature reuse through pre-normalization, channel expansion, depthwise temporal filtering, and gated residual learning.

Given the input $\mathbf{F}_{i-1}\in\mathbb{R}^{T' \times C}$ of the $i$-th TEAB, where $C$ is the channel dimension, we first apply layer normalization and pointwise expansion:
\begin{equation}
\mathbf{H}_{i}
=
\delta\left(
\mathrm{PWConv}_{e}\big(
\mathrm{LN}(\mathbf{F}_{i-1})
\big)
\right),
\label{eq:teab_h}
\end{equation}
where $\mathrm{PWConv}_{e}(\cdot)$ denotes pointwise convolution with expansion factor $e$, expanding the channel dimension from $C$ to $eC$.

Temporal modeling is then performed by depthwise convolution:
\begin{equation}
\mathbf{G}_{i}
=
\mathrm{Drop}\left(
\mathrm{BN}\left(
\delta\left(
\mathrm{DWConv}(\mathbf{H}_{i})
\right)\right)\right),
\label{eq:teab_g}
\end{equation}
followed by a sigmoid gate:
\begin{equation}
\mathbf{A}_{i}
=
\sigma(\mathbf{G}_{i}),
\label{eq:teab_a}
\end{equation}
and gated residual refinement:
\begin{equation}
\mathbf{F}_{i}
=
\mathbf{F}_{i-1}
+
\mathbf{A}_{i}\odot\mathbf{F}_{i-1}.
\label{eq:teab_out}
\end{equation}

This formulation preserves the shortcut path while adaptively emphasizing temporally salient emotion cues. For bidirectional modeling, TEAB is recursively applied in both directions:
\begin{equation}
\overrightarrow{\mathbf{F}}_{i}
=
\mathcal{T}_{i}(\overrightarrow{\mathbf{F}}_{i-1}), 
\qquad
\overleftarrow{\mathbf{F}}_{i}
=
\mathcal{T}_{i}(\overleftarrow{\mathbf{F}}_{i-1}).
\label{eq:teab_bi}
\end{equation}

\subsection{Weighted Bidirectional Fusion}
\label{subsec:wbf}

After obtaining forward and reverse features at each level, we fuse them using a \textbf{weighted bidirectional fusion} (BiF) module instead of the direct summation in TIM-Net. The goal is to adaptively balance the two temporal directions and the contributions of different levels.


Unlike conventional layer-specific weighting, we introduce two global learnable parameters, $a$ and $b$, which are shared across all levels to scale the forward and reverse temporal directions, respectively. Specifically, at level $i$, the bidirectional features are fused as follows:
\begin{equation}
\mathbf{f}_{i} 
= 
\mathrm{GAP} \big(
a\overrightarrow{\mathbf{F}}{i} + b\overleftarrow{\mathbf{F}}_{i}
\big),
\label{eq:wbf_fi}
\end{equation}
Finally, the fused features from all levels are aggregated to form the final utterance-level representation:
\begin{equation}
\mathbf{f} = \sum_{i=1}^{n}\lambda_{i}\mathbf{f}_{i},
\label{eq:wbf_final}
\end{equation}
where $\lambda_i$ is a learnable weight parameter specific to the $i$-th level. This two-stage weighting mechanism allows the model to flexibly balance both the temporal directions (via global shared weights $a$ and $b$) and the hierarchical levels (via level-specific weights $\lambda_i$), yielding a more expressive representation than the original TIM-Net.

\section{Experiments}

\begin{table*}[t]
\centering
\caption{Comparison of UA and F1 (\%) across five SER benchmarks. 
Model complexity is reported as the number of parameters(M) and FLOPs(G), computed on 5-second input utterances. 
The best and second-best results in each column are highlighted in \textbf{bold} and \underline{underlined}, respectively. $R$ denotes the subsampling rate. This convention is followed in all subsequent tables.}
\label{tab:6datasets_avg}
\small
\setlength{\tabcolsep}{3.8pt}
\begin{tabular}{l c cc cc cc cc cc cc}
\toprule
\multirow{2}{*}{\textbf{Model}} 
& \multirow{2}{*}{\textbf{Comp. (M/G)}} 
& \multicolumn{2}{c}{\textbf{EMOVO}} 
& \multicolumn{2}{c}{\textbf{IEMOCAP}} 
& \multicolumn{2}{c}{\textbf{RAVDESS}} 
& \multicolumn{2}{c}{\textbf{CREMA-D}} 
& \multicolumn{2}{c}{\textbf{MELD}} 
& \multicolumn{2}{c}{\textbf{Avg}} \\
\cmidrule(lr){3-4} \cmidrule(lr){5-6}
\cmidrule(lr){7-8} \cmidrule(lr){9-10}
\cmidrule(lr){11-12} \cmidrule(lr){13-14}
& & UA & F1 & UA & F1 & UA & F1 & UA & F1 & UA & F1 & UA & F1 \\
\midrule

wav2vec 2.0 base \cite{baevski2020wav2vec}
& $\sim$95\,/\,33.53
& \underline{31.07} & \underline{27.24}
& \underline{58.27} & \textbf{57.83}
& 54.33 & \underline{53.99}
& 61.95 & 61.75
& 20.06 & \textbf{20.04}
& 45.14 & 44.17 \\

TIM-Net \cite{ye2023temporal}
& $\sim$0.12\,/\,0.05
& 21.94 & 15.71
& 54.91 & 52.57
& 50.67 & 49.84
& 66.00 & 65.68
& 17.60 & 15.37
& 42.22 & 39.83 \\

MS-SENet \cite{10447232}
& $\sim$0.15\,/\,0.13
& 22.11 & 13.77
& 55.48 & 54.07
& 54.62 & 53.77
& 65.10 & 64.74
& 17.55 & 14.73
& 42.97 & 40.22 \\

\midrule

\textbf{SETEAB ($R=2$)}
& $\sim$0.5\,/\,0.12
& \textbf{35.20} & \textbf{31.61}
& 58.07 & 56.26
& \underline{54.95} & 53.26
& \textbf{67.29} & \textbf{67.04}
& \textbf{22.97} & 17.51
& \textbf{47.69} & \underline{45.14} \\

\textbf{SETEAB ($R=4$)}
& $\sim$0.4\,/\,0.06
& 30.27 & 24.94
& \textbf{58.94} & \underline{57.12}
& \textbf{59.57} & \textbf{59.64}
& \underline{66.73} & \underline{66.55}
& \underline{22.54} & \underline{17.89}
& \underline{47.61} & \textbf{45.23} \\

\bottomrule
\end{tabular}
\end{table*}

\subsection{Experimental Setup}

\subsubsection{Datasets and evaluation metrics}


To evaluate the proposed framework, we adopt the EmoBox benchmark protocol \cite{ma24b_interspeech}, ensuring standardized data partitions and reproducible comparisons. For intra-corpus experiments, we utilize EMOVO \cite{costantini-etal-2014-emovo}, IEMOCAP \cite{dataset_IEMOCAP}, RAVDESS \cite{dataset_RAVDESS}, MELD \cite{dataset_MELD}, and CREMA-D \cite{dataset_cremad}, reporting unweighted accuracy (UA) and macro-F1 scores to reflect class-balanced performance. For cross-corpus experiments, we follow the same benchmark and evaluate on IEMOCAP, RAVDESS, MELD, and SAVEE \cite{dataset_SAVEE}, reporting weighted accuracy (WA) to assess generalization under domain shifts and real-world conditions.
\subsubsection{Implementation details}

In our experiments, all audio signals are first resampled to 16~kHz. We then extract 80-dimensional log-Mel spectrograms using a 25~ms analysis window and a 10~ms frame shift, followed by cepstral mean normalization. To improve generalization and reduce overfitting, we apply stochastic data augmentation with probability $P = 0.2$, including random time shifting ($\pm 5$ frames), pitch perturbation ($\pm 2$ semitones), joint speed-pitch scaling (0.8$\times$ to 1.2$\times$), and time stretching with a factor of 0.8. In addition, SpecAugment \cite{park19e_interspeech} is applied to the spectral features during training.

All models are optimized using Adam \cite{adam_opt_2014} with an initial learning rate of $\alpha = 0.001$, momentum coefficients $\beta_1 = 0.93$ and $\beta_2 = 0.98$, and a weight decay of $10^{-6}$. The learning rate is decayed by a factor of 0.98 at each epoch. To further mitigate overfitting, we apply label smoothing with a factor of 0.1 and early stopping with a patience of 20 epochs. The cross-entropy objective is used for training. 

For the $i$-th TEAB, the channel dimension is fixed at $C=64$. Each block contains a pointwise convolution (expansion ratio $e=4$, kernel size $k=1$), followed by a depthwise convolution with a kernel size of 3. We use a dropout rate of 0.1 and set the dilation factor to $2^{i-1}$. The number of TEABs in each direction is fixed at $n=8$. The batch size is set to 16 for MELD and IEMOCAP, and 32 for the remaining datasets, due to the hardware constraints of a workstation equipped with an Intel Core i7-12800 CPU, 32~GB RAM, and an RTX A1000 GPU (4~GB VRAM).

\subsection{Results and Analysis}

\subsubsection{Intra-corpus SER Results}
Table~\ref{tab:6datasets_avg} compares two variants of our model with TIM-Net and two recent SER baselines (wav2vec~2.0 base and MS-SENet) on five benchmark datasets to assess the accuracy-efficiency trade-off. Overall, our variants achieve the best performance: \emph{SETEAB} ($R=2$) yields the highest average UA (47.69\%), while \emph{SETEAB} ($R=4$) attains the best average F1 (45.23\%), with each variant ranking second on the complementary metric. Both consistently outperform TIM-Net and MS-SENet across both UA and F1. Despite these accuracy gains, our models remain lightweight. Compared to wav2vec~2.0 base (95M parameters, 33.53GFLOPs), our variants require only 0.4--0.5M parameters and 0.06--0.12GFLOPs, demonstrating a substantially improved accuracy-efficiency trade-off.

\begin{figure}[t]
    \centering
    \includegraphics[width=\columnwidth]{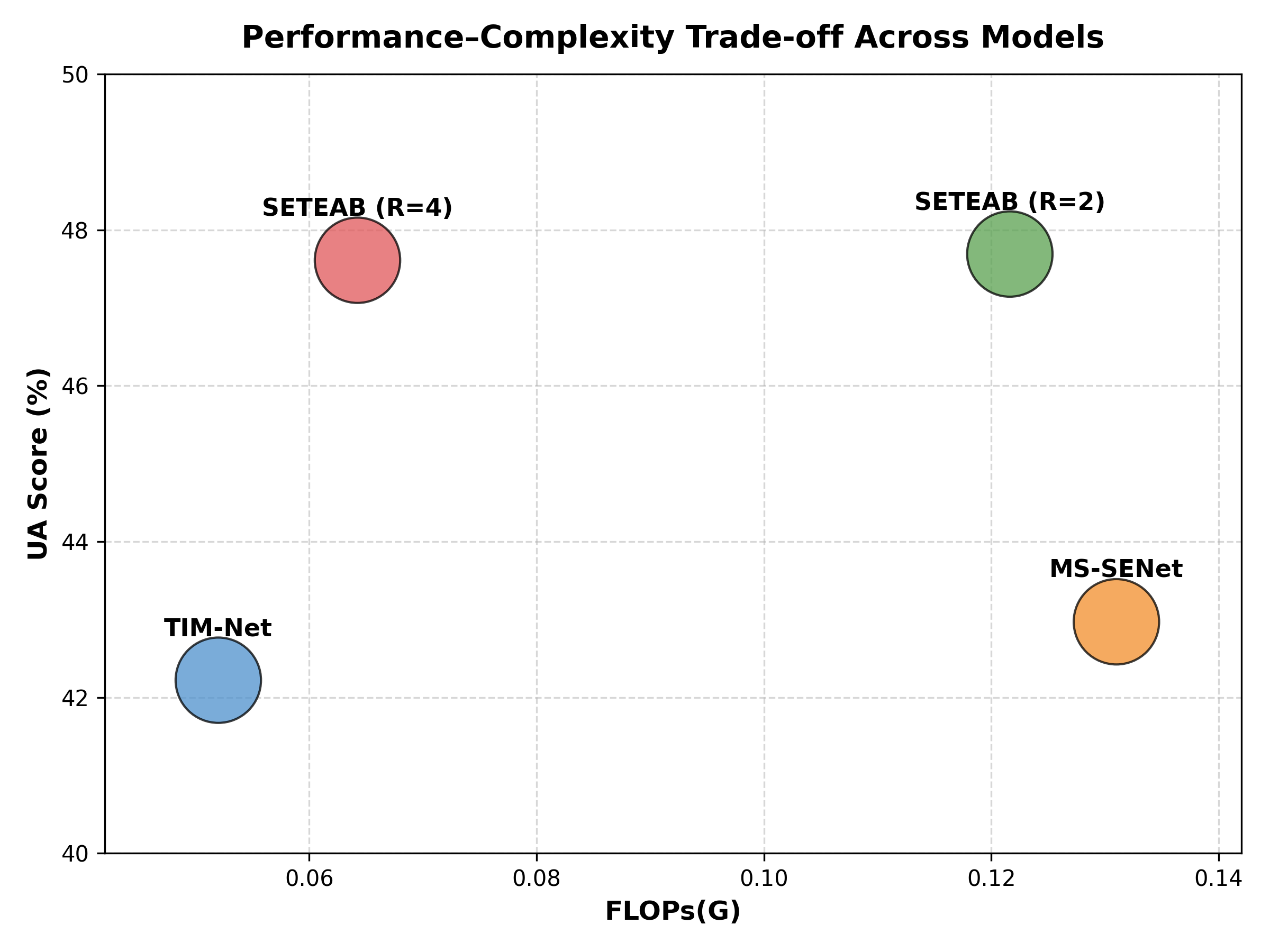}
    \caption{Comparison of performance–efficiency trade-off in terms of FLOPs(G) and UA\% across baseline and our models.}
    \label{fig:comparision}
\end{figure}

Fig.~\ref{fig:comparision} illustrates the trade-off between computational cost and recognition performance among the compared models. The proposed variants achieve a superior balance compared to the baselines. In particular, \emph{SETEAB} ($R=4$) delivers very strong recognition performance while requiring only 0.06~GFLOPs, which is substantially lower than both MS-SENet and \emph{SETEAB} ($R=2$). Although \emph{SETEAB} ($R=2$) achieves the highest UA, the $R=4$ variant is especially attractive because it preserves competitive accuracy at a much lower computational cost, making it a highly efficient choice for practical SER systems.

\begin{table}[t]
\centering
\caption{Ablation study of the proposed model variants on SER datasets, reported in terms of average UA (\%) and F1 (\%).}
\label{tab:ablation_ours}
\small

\resizebox{\columnwidth}{!}{
\begin{tabular}{lccc}
\toprule
\multicolumn{1}{c}{\textbf{Setting}} & \textbf{FLOPs(G)} & $\mathbf{UA_{avg}}$ & $\mathbf{F1_{avg}}$  \\
\midrule

DW-Sub ($R=2$) + BiF + SE-Res2 & 0.12 & \textbf{47.69} & \underline{45.14} \\
DW-Sub ($R=4$) + BiF + SE-Res2 & 0.06 & \underline{47.61} & \textbf{45.23} \\

\midrule
BiF & 0.14 & 45.89 & 43.68 \\
BiF + SE-Res2 & 0.16 & 46.22 & 43.20 \\
Bi-DW-Sub + BiF + SE-Res2 & 0.17 & 46.62 & 43.96 \\

\bottomrule
\end{tabular}
}
\end{table}


Table~\ref{tab:ablation_ours} shows that each proposed component contributes positively to SER performance. Starting from the BiF baseline, adding SE-Res2 improves feature representation. Introducing depthwise subsampling (DW-Sub), including the variant operating in forward and backward directions (Bi-DW-Sub), further improves the average UA and F1 scores. The full model achieves the best overall performance, confirming the effectiveness of the proposed components.

\subsubsection{Cross-corpus SER Results}

\begin{table}[ht]
\centering
\caption{Cross-dataset evaluation (WA\%) across four datasets (IEMOCAP (I), MELD (M), RAVDESS (R), SAVEE (S)).}
\label{tab:cross_dataset}
\small
\resizebox{\linewidth}{!}{
\begin{tabular}{c|cccc|cccc}
\toprule
\multirow{2}{*}{\textbf{Test}} 
& \multicolumn{8}{c}{\textbf{Training Set}} \\ \cline{2-9}

& \multicolumn{4}{c|}{\textbf{wav2vec 2.0 base} } 
& \multicolumn{4}{c}{\textbf{SETEAB (ours)}} \\ \cline{2-9}
& \textbf{I} & \textbf{M} & \textbf{R} & \textbf{S}
& \textbf{I} & \textbf{M} & \textbf{R} & \textbf{S} \\
\midrule

\textbf{I} 
& $\backslash$ & 29.78 & 18.25 & 28.84 
& $\backslash$ & \textbf{47.08} & \textbf{37.50} & 30.42 \\

\textbf{M} 
& 22.50 & $\backslash$ & \textbf{31.39} & 35.24
& \textbf{35.00} & $\backslash$ & 27.08 & 30.00 \\

\textbf{R} 
& 27.15 & 23.20 & $\backslash$ & 33.77
& \textbf{48.33} & \textbf{44.17} & $\backslash$ & \textbf{47.08} \\

\textbf{S} 
& 31.34 & \textbf{29.19} & 21.36 & $\backslash$
& \textbf{44.17} & 27.50 & 32.08 & $\backslash$ \\

\midrule
& \multicolumn{4}{c|}{\textbf{TIM-Net}} 
& \multicolumn{4}{c}{\textbf{MS-SENet}} \\ \cline{2-9}

\textbf{I} 
& $\backslash$ & 33.33 & 35.83 & 34.58 
& $\backslash$ & 35.42 & 35.83 & \textbf{39.17} \\

\textbf{M} 
& 25.00 & $\backslash$ & 26.67 & 30.83 
& 26.67 & $\backslash$ & 29.58 & \textbf{35.42} \\

\textbf{R} 
& 30.83 & 37.92 & $\backslash$ & 34.58 
& 34.58 & 40.00 & $\backslash$ & 34.58 \\

\textbf{S} 
& 35.83 & 22.50 & \textbf{38.75} & $\backslash$
& 37.92 & 27.08 & 35.42 & $\backslash$ \\

\bottomrule
\end{tabular}
}
\end{table}

Table~\ref{tab:cross_dataset} shows that the proposed \textbf{SETEAB} achieves stronger cross-dataset generalization than the baselines in most train-test settings. In particular, it obtains the best results in \textbf{7 out of 12} cross-corpus evaluations, outperforming wav2vec~2.0 base, TIM-Net, and MS-SENet overall. The gains are especially clear when testing on \textbf{RAVDESS}, where SETEAB consistently delivers the highest accuracy across all three training sources. These results indicate that the proposed combination of depthwise subsampling, SE-Res2Block, TEAB, and weighted bidirectional fusion learns more transferable emotional representations. Overall, SETEAB provides a better balance between recognition accuracy and cross-corpus robustness, which is crucial for practical SER systems.

\begin{table}[ht]
\centering
\caption{Overall cross-dataset WA (\%) of different models, reported as mean $\pm$ standard deviation.}
\label{tab:wa_mean_std}
\small
\begin{tabular}{lc}
\toprule
\textbf{Model name} & \textbf{WA} \\
\midrule
wav2vec 2.0 base & 27.67 $\pm$ 5.27 \\
TIM-Net          & 32.22 $\pm$ 5.18 \\
MS-SENet         & 34.31 $\pm$ 4.35 \\
\midrule
\textbf{SETEAB (ours)}             & \textbf{37.53 $\pm$ 8.20} \\

\bottomrule
\end{tabular}
\end{table}

Table~\ref{tab:wa_mean_std} summarizes the overall cross-corpus WA, where \textbf{SETEAB} achieves the best mean performance at \textbf{37.53\%}, outperforming wav2vec~2.0 base, TIM-Net, and MS-SENet. The consistent improvement in mean WA indicates that our design learns \textbf{more transferable emotional representations} under dataset shifts. This is particularly important for SER in the wild, where training and deployment domains rarely match. Although the variance is larger, the higher mean suggests stronger robustness across diverse cross-corpus conditions. Overall, SETEAB provides a clear generalization advantage in cross-dataset evaluations.

\section{Conclusion}

This paper presented \textbf{SETEAB}, a lightweight architecture for speech emotion recognition. The proposed model improves TIM-Net through four key designs: \textbf{depthwise convolution subsampling} for model compression, \textbf{Squeeze-and-Excitation} for channel-wise refinement, a novel \textbf{TEAB} for enhanced temporal modeling, and \textbf{weighted bidirectional fusion} for adaptive integration of temporal cues. Experimental results demonstrate that SETEAB achieves a strong accuracy-efficiency trade-off, with a compact model size, high recognition performance, and stable cross-corpus generalization. These findings confirm the effectiveness and generalization ability of the proposed approach for robust SER.

\section{Acknowledgments}
This research was funded by University of Science, VNU-HCM under grant number CNTT 2025-20.

\section{Generative AI Use Disclosure}
During the preparation of this manuscript, the authors used AI-assisted tools to improve the readability and clarity of the text; however, the authors take full responsibility for the content, accuracy, and integrity of the paper.

\bibliographystyle{IEEEtran}
\bibliography{mybib}

\end{document}